\documentclass[aps,prl,superscriptaddress]{revtex4}
\usepackage{graphicx}



\begin{document}

\title{Coherent scattering from silicon monocrystal surface}

\author{Fr\'ed\'eric Livet}

\affiliation{SIMaP, Grenoble-INP, CNRS, UJF
\\
 BP 75, 38402 Saint Martin D'H\`eres, France}
\email{frederic.livet@simap.grenoble-inp.fr}

\author{Guillaume Beutier}

\affiliation{Diamond Light Source\\
Harwell Science \& Innovation Campus, OX11 0DE, United Kingdom
}

\author{Marc de Boissieu}

\affiliation{SIMaP, Grenoble-INP, CNRS, UJF
\\
 BP 75, 38402 Saint Martin D'H\`eres, France}

\author{Sylvain Ravy}
\author{Fr\'ed\'eric Picca}

\affiliation{Synchrotron Soleil\\
L'Orme des Merisiers, Saint-Aubin, BP 48, 91192 GIF-sur-YVETTE
France}

\author{David Le Bolloc'h}
\author{Vincent Jacques}

\affiliation{LPS, CNRS, Universit\'e Paris Sud\\
B\^atiment 510, 91405 Orsay, France
}

\date{\today}

\begin{abstract}
Using coherent x-ray scattering, we evidenced atomic step roughness 
at the [111] vicinal surface of a silicon monocrystal of 0.05$^{\circ}$ 
miscut. Close to the ($\frac{1}{2} \frac{1}{2} \frac{1}{2}$) anti-Bragg 
position of the reciprocal space which is particularly sensitive 
to the [111] surface, the truncation rod exhibits a contrasted 
speckle pattern that merges into a single peak closer to the (111) 
Bragg peak of the bulk. The elongated shape of the speckles along the
[111] direction confirms the monoatomic step sensibility of the technique. 
This experiment opens the way towards studies of step dynamics on 
crystalline surfaces.

\end{abstract}

\maketitle

The surface of crystals is extensively studied because it is
the region of the sample where elements are aggregated from liquid,
vapor or solid phases. Numerous methods have been developed for
the observation of surface morphology. Optical or Scanning
Electron Microscopy methods are the simplest \cite{Chata93}. Low energy
electron reflection microscopy \cite{Mulle05,Bauer94} and near field methods 
(AFM, MFM, STM) \cite{Surne97,Barth07}, are now classical 
methods for the study of the static surface configuration.

An important subject of the surface study is the direct 
observation of the dynamics of the surface fluctuations connected 
to surface step movement \cite{Kuipe95,Giese98,Dough03}, to
change in crystal shape by nucleation and/or annihilation of 
steps at face edges \cite{Thuer01} or to the fluctuations of
chemically inhomogeneous surfaces \cite{lyubi02}. The
main available experimental method for the observation of
surface dynamics is time-resolved STM, where the movement
of an individual step is observed by successive
linear scans across a reduce segment. In these experiments,
the evolution of the step position is recorded , and from 
a series of measurements, a statistical study provides information
on the dynamics of surface evolution \cite{Dough03,LeGof03}.
This method provides measurements of the step dynamics
for times ranging between 0.1 and a few tens of seconds and for 
distances between 0.1 and a few tens of nanometers. These very local 
measurements provide accurate knowledge of the atomic surface 
properties, and the various atomic models of the dynamics 
can be discussed (see Giesen \cite{Giese01}). 

X-ray diffraction has also been proved a very useful method for 
surface observation \cite{Robin86}. The regions of the reciprocal 
space where valuable information about the surface 
structure can be obtained are the ``truncation rods'' (TR). TR are 
satellite streaks perpendicular to the surface originating from Bragg 
peaks, and by analogy, specular reflectivity can be considered as the 
satellite of the (000) reciprocal position. The longitudinal variations 
of TR intensity have been extensively used for the observation of the 
change in atomic position \cite{Robin92} and in atomic composition in 
the vicinity of the surface \cite{Robin88,Robac03} and of surface 
roughness \cite{Munkh97}.
  
Moreover, the transverse intensity of the TR corresponds to
the Fourier transform of the surface bidimensional array. 
Qualitatively, the relative distance ($\Delta Q\simeq2\pi/e$)
of the reciprocal point of the TR from the nearest Bragg peak 
samples a thickness $e$ under the surface. The special
position intermediate between two Bragg peaks essentially 
corresponds to surface defects of monoatomic thickness (steps,
monoatomic layers..) and is called the ``anti-Bragg'' (AB) region.
At this point, perfectly flat diffracting planes scatter with
relative phase shifts of $\pi$, cancelling each other and only the 
scattering of the cut-off introduced by the surface is observed.

In this region, the per sample atom cross section is
very low, and the Bragg to AB cross section ratio roughly
corresponds to the square of the ratio between the 
number of volume atoms and the number of surface 
atoms in the coherence volume \cite{Robin86}. The AB intensity 
is also proportional to the number of irradiated surface atoms, and 
measurements are often carried out in asymmetric geometry, with 
a low incident angle.

In this letter, we show that the TR tranverse intensity in the
conditions of coherent diffraction can provide a new method
for the observation of steps at the surface of crystals and that
this method can be extended to dynamical surface studies.

\begin{figure}[tbp]
\includegraphics[width=6.cm]{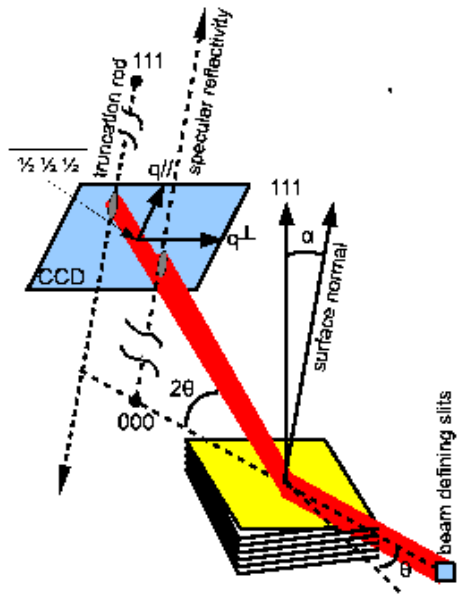}
\caption{Description of the experiment. The miscut angle is 
$\alpha$. The CCD selects a small plane of the reciprocal
lattice and intercepts the (111) direction with an angle
$\theta$. The direction $q_{\perp}$ is horizontal 
and $q_{\parallel}$ is in the vertical diffracting plane}
\label{fig:setup}
\end{figure}

Figure~\ref{fig:setup} gives a rapid scheme of the setup used for
this type of experiment. Coherent diffraction from a sample can 
be observed if the irradiated volume has the size of the coherence 
volume. A good transverse coherence length is obtained by using 
suitable slits, in order that the product of the beam size at 
sample $\phi$ by the beam divergence $\epsilon$ is of the order 
of the wavelength $\lambda$ \cite{Livet07-3}.
The longitudinal coherence length $\Lambda_l$ is fixed by the beam 
monochromaticity $\Lambda_l=\lambda^2 / 2\Delta \lambda$, and the 
path-length difference $\Delta\cal L$ of the beam in the sample 
must not exceed this limit. This latter condition is 
contradictory with the asymmetric geometry. As
scattering far enough from Bragg peaks is discussed here
for the observation of surfaces, only interferences between 
surface atoms are relevant. For flat surfaces, with a miscut angle 
$\alpha$ (see Fig.~\ref{fig:setup}), the pathlength differences in 
asymmetric conditions can be written:

\begin{equation}
\Delta {\cal L} \simeq 2\phi \times \alpha_{\parallel} \leq \Lambda_l
\label{equ:coh_vert}
\end{equation}
where $\alpha_{\parallel}$ is the asymmetry angle (assumed small) in the 
diffraction plane (see Fig.~\ref{fig:setup}). This means that 
$\alpha_{\parallel} \leq \Lambda/(2\phi)$.
For symmetric diffraction geometry, one has also to take account
of the angle $\alpha_{\perp}$ between $\vec Q$ and the surface 
normal:

\begin{equation}
2 \phi \tan(\alpha_{\perp}) \sin(\theta) \leq \Lambda_l
\label{equ:coh_hor}
\end{equation}

and $\tan{\alpha_{\perp}} \leq \Lambda_l/(2 \phi \sin(\theta))$ provides
a limit for the directions of $\vec Q$ that can be explored.

For the observation of speckles in the truncation rod, we have chosen
to test a silicon crystal with a very small misorientation of the 
[111] axis relative to the surface normal (a vicinal surface with an
$\alpha\simeq 0.05 ^{\circ}$ miscut). Silicon gives perfect single 
crystals, and the crystal studied here had been carefully polished by 
the ESRF staff for monochromator use.
Laboratory x-ray reflectivity measurements showed a few Angstroms  
rugosity of the surface and a very thin (less than one nanometer 
on average) layer of silicon oxide. Coherent x-ray scattering 
experiments were performed at the CRISTAL beamline of the 
synchrotron SOLEIL. The 1.772 \AA\ (7~keV) wavelength was selected with a 
Si$_{111}$ double crystal monochromator and mirrors were used in order to 
suppress harmonics.
One can estimate here $\Lambda_l \simeq 0.6 \mu$m. Transverse coherence
is achieved by combining two sets of square slits, one 13 m upward, 
with a 100 to 250~$\mu$m aperture and one close (0.15~m) to the sample, 
with 10 to 20~$\mu$m apertures. By this setup, we can adjust the balance 
between intensity and coherence. Speckles of reasonable contrast were 
observed with the maximum aperture, the beam intensity being in the 
$2.10^8$~ph/s range after the slits.

The sample was in air and positioned precisely at the center of the 
kappa goniometer of the beamline. The scattering plane
was vertical. Scattering was measured with a back illuminated 
CCD from Andor Technologies (1024$\times$1024 pixels of 13~$\mu$m size) 
located at 2.2~m. A dedicated  program was used 
for individual photon extraction from each frame \cite{Livet00,Beuti08}. 
For low intensity measurements, this eliminates the CCD electronic 
noise. Static images were obtained from the accumulation of at least 
100 frames. With the fast (2.5~Mhz) converter, good quality images 
could be obtained in one minute for large enough intensity.

\begin{figure}[tbp]
\includegraphics[clip,width=6.cm]{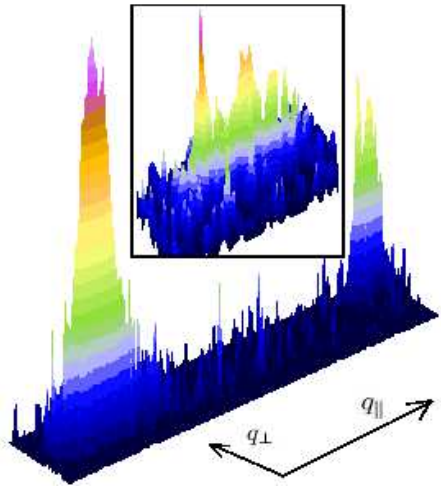}
\caption{Image of the observed scattering in the vicinity of the 
($\frac{1}{2} \frac{1}{2} \frac{1}{2}$) AB position.
Log units, maximum intensity for the TR is 50~ph/pixel for a 2,500~s 
acquisition time. The insert shows the detailed shape of the right side 
peak in linear units, showing speckles though a maximum of 15~counts in 
2,500 s. Air scattering is less than one count.
}
\label{fig:twopeaks}
\end{figure}

Diffraction was studied in the vicinity of the ($hhh$) positions, $h$
varying from $1/2$ to 1. In Fig.~\ref{fig:setup} the 
reciprical lattice plane observed for $h=1/2$ (AB for the
(000) and (111) reciprocal lattice points) is schematized. This 
plane intercepts the specular and the truncation rods. 
Fig.~\ref{fig:twopeaks} shows the typical image of the diffracted 
intensity around the AB position. 

Although the area of the reciprocal lattice covered by our detector is 
very small, we observe here two peaks, with $n_{\parallel}=550$~pixels 
and $n_{\perp }=50$~pixels spacings in the $q_{\perp}$ and 
$q_{\parallel}$ directions respectively. The higher intensity
peak (about 2.~ph/s total intensity) corresponds 
to the truncation rod and the lower intensity one (about 0.3~ph/s 
total intensity) to the surface reflectivity. 
Both peaks are roughly aligned in the vertical direction of the 
detector, essentially because the TR is intercepted with a small 
$\theta$ angle (8.$^{\circ}$). 
The distance between these two peaks provide a very precise 
estimate of the miscut  $\alpha$. In the configuration of 
Fig.~\ref{fig:twopeaks}, we obtain:
$\alpha_{\parallel}=n_{\parallel}13.10^{-6}/(4\times 2.2)=0.047^{\circ}$ and 
$\alpha_{\perp}=n_{\perp}13.10^{-6}/(4\times2.2\times \sin(\theta))=0.030^{\circ}$ 
which gives the miscut angle: 
$\alpha=\sqrt{\alpha_{\parallel}^2+\alpha_{\perp}^2}=0.056^{\circ}$.

The lower intensity peak, detailed in the insert of Fig.~\ref{fig:twopeaks}, 
is a large $|\vec{Q}|$ measurement of the specular reflectivity and its 
intensity is connected to surface rugosity. This involves the silicon 
crystal rugosity and the thin amorphous silicon oxide layer. In 
our laboratory experiment (incoherent scattering) we have observed that 
the contribution of this oxide layer lowers the pure silicon 
reflectivity by a factor ten at this Q-value. 

The synchrotron experiment described here differs from incoherent 
scattering experiments: in a laboratory experiment, the irradiated 
region of the sample ranges in millimeters with a few $\mu$m 
coherence length, while here, all the illuminated 
140~$\mu$m$\times 20 \mu$m region scatters coherently 
\footnote{in this case, the maximum opening of the slits was used, 
this gives a $2.10^8$~ph/s intensity, with a lower coherence contrast}.
This can be checked from the insert of Fig.~\ref{fig:twopeaks} 
(linear plot), where we observe that this peak is elongated in the
direction of q$_{\parallel}$, and that its profile has a speckle 
structure corresponding to surface rougheness. This typical behavior 
was observed for much smaller $\theta$ angles by Robinson et 
al.~\cite{Robin98}, with a much larger elongation.

\begin{figure}
\begin{tabular}{cc}
\includegraphics[clip,width=3.6cm] {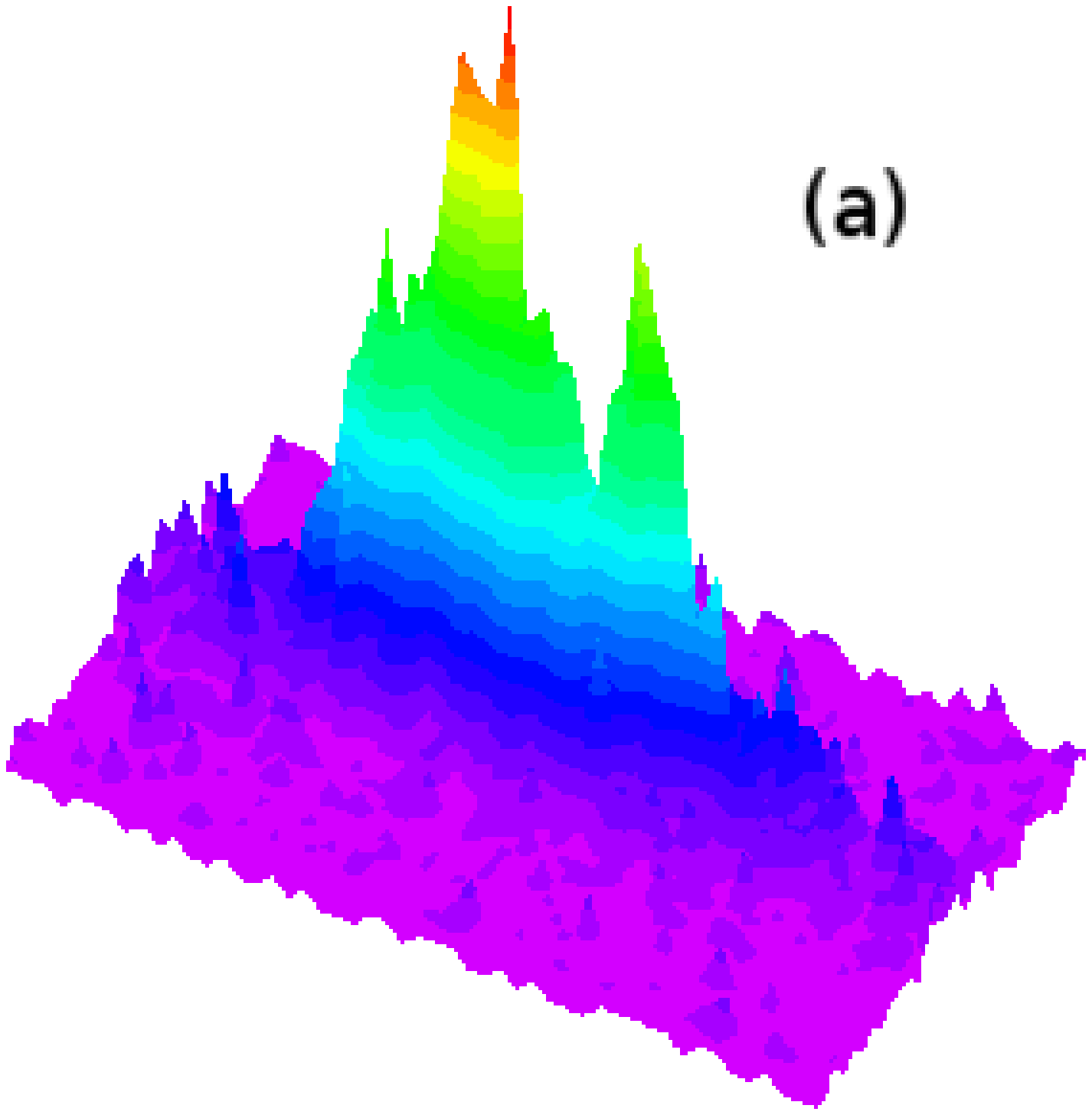}
	&\includegraphics[clip,width=3.6cm] {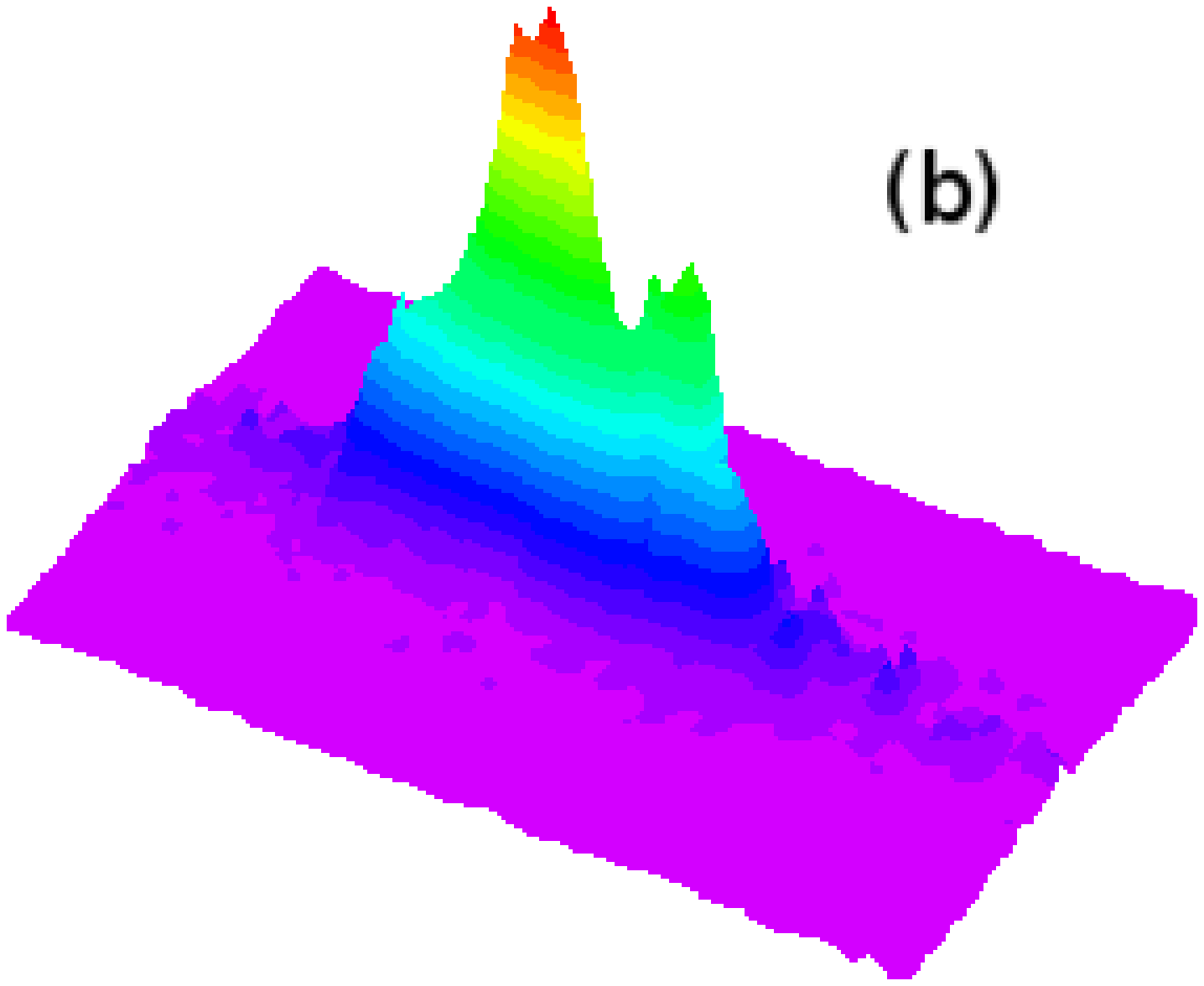}\\
\includegraphics[clip,width=3.6cm] {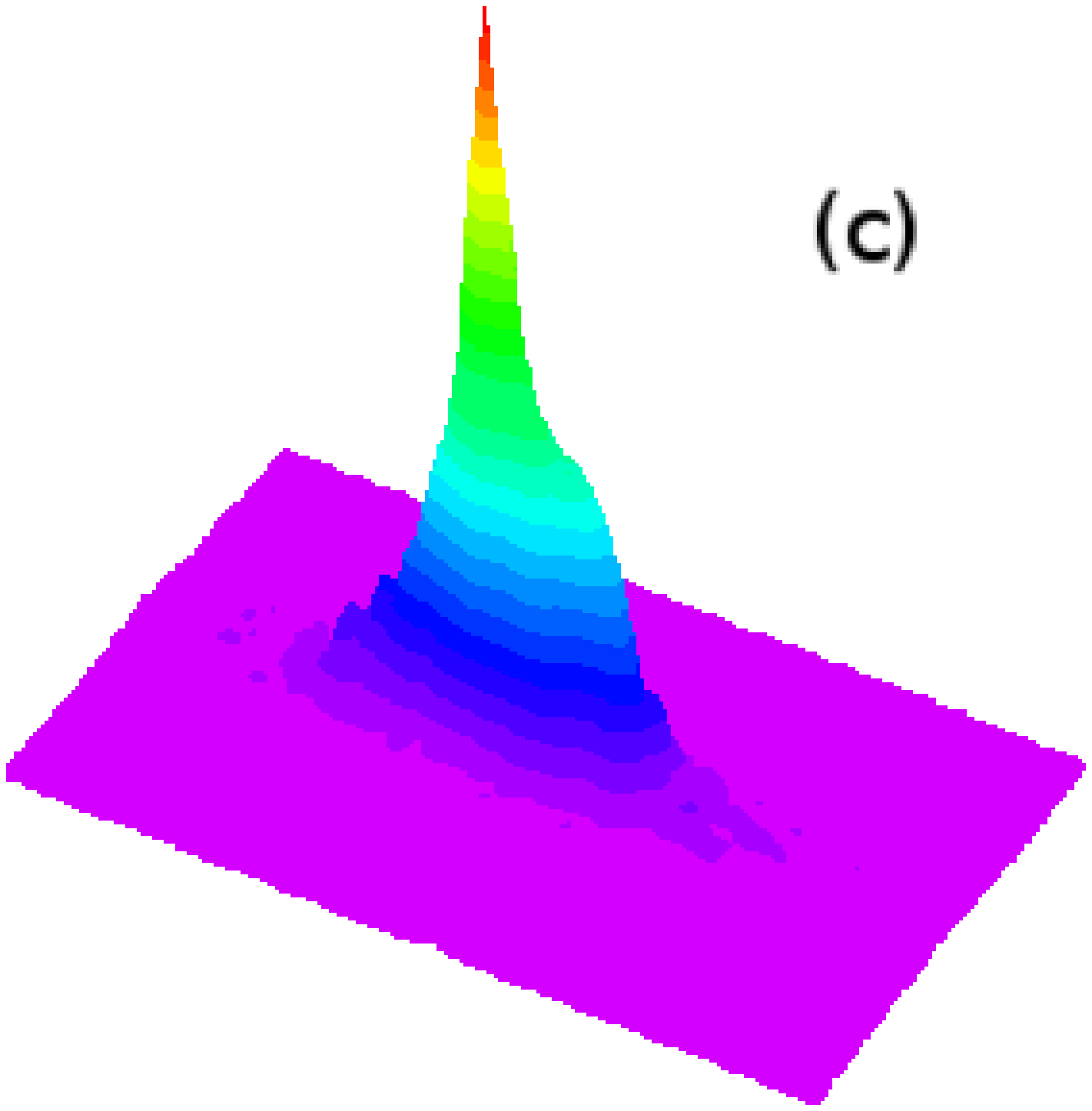}
	&\includegraphics[clip,width=3.6cm] {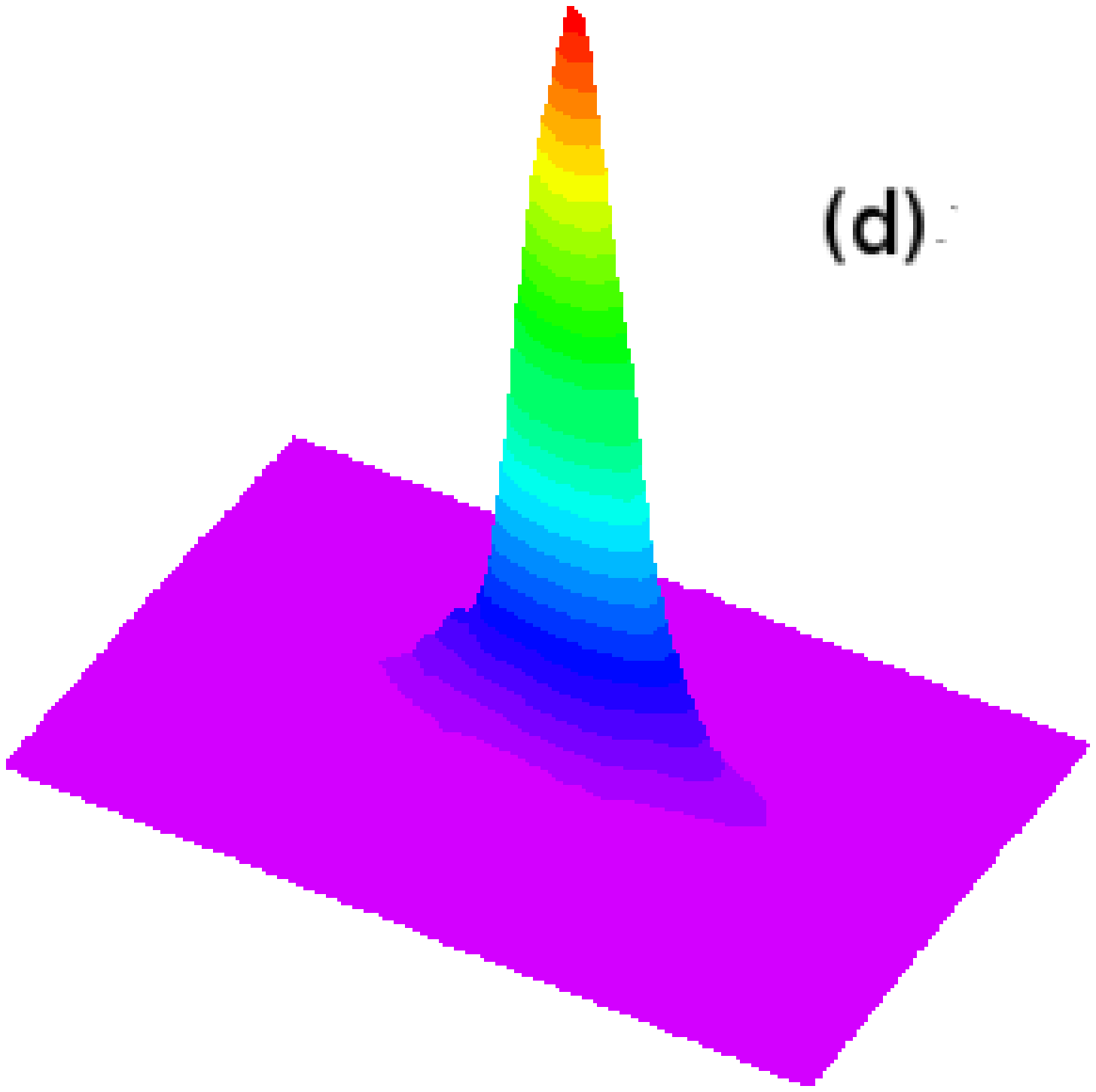}\\
\includegraphics[clip,width=3.6cm] {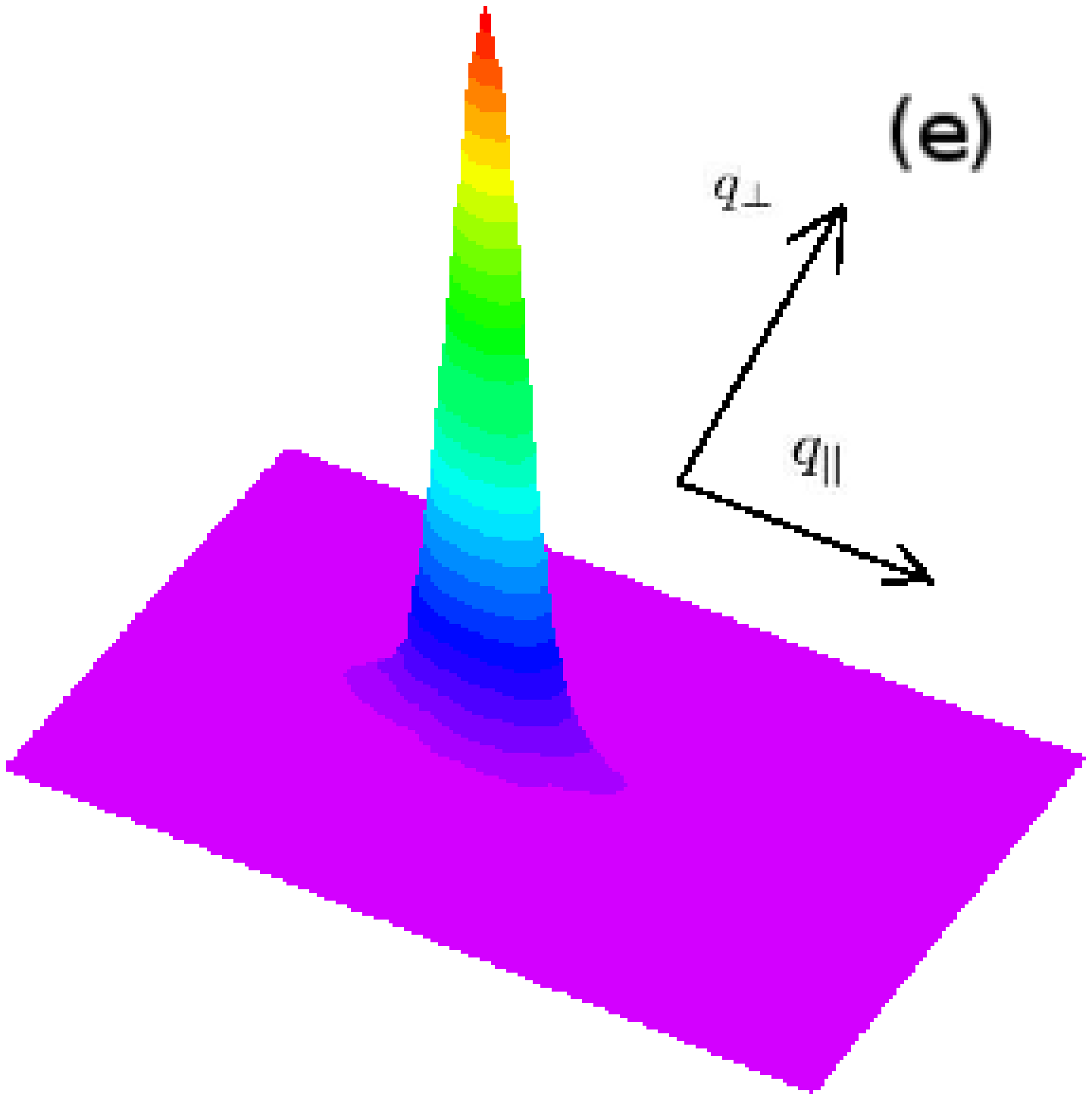}
	&\includegraphics[clip,width=3.6cm] {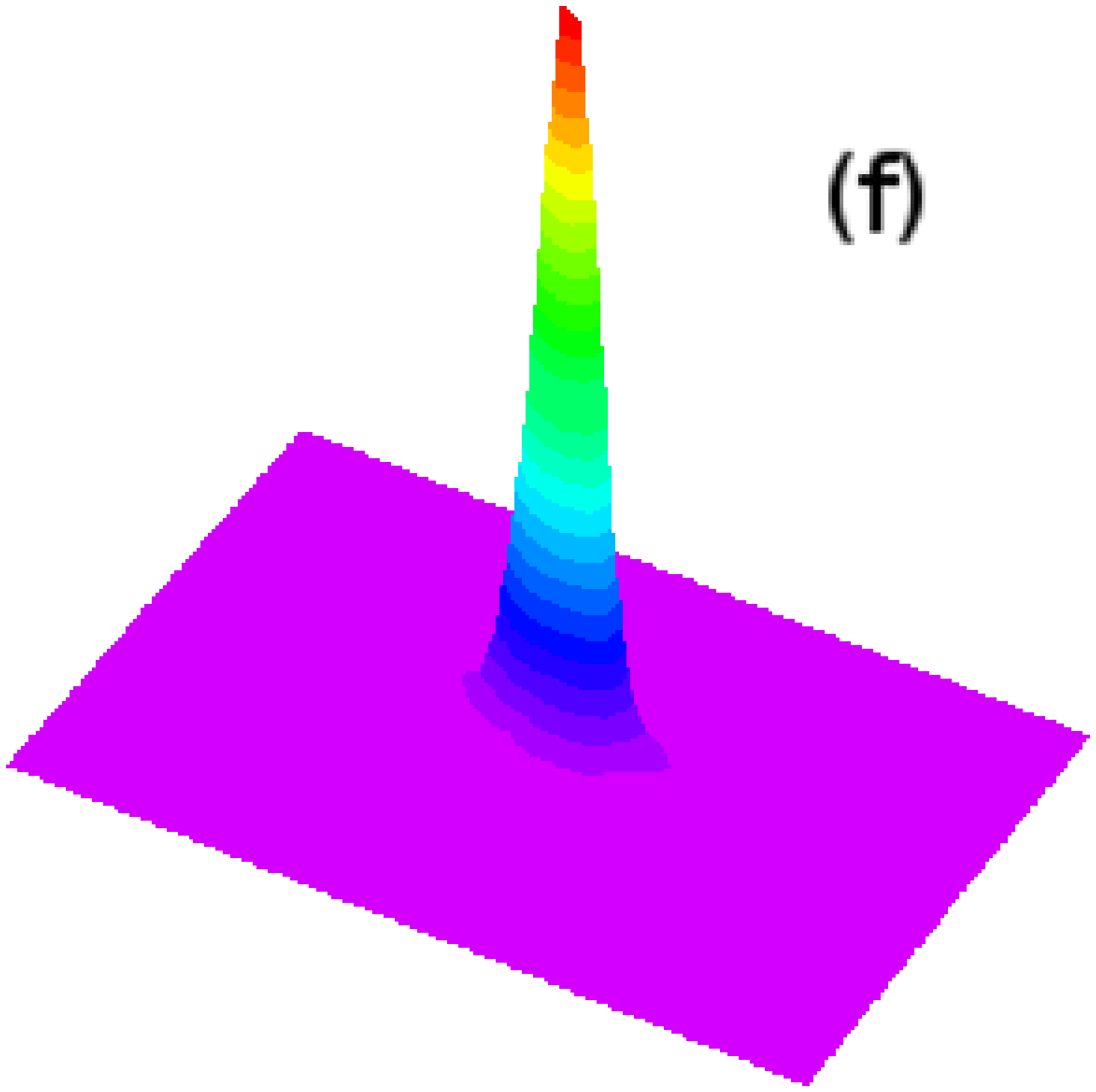}\\

\end{tabular}
\caption {The shape of the truncation rod for various (h,h,h) vectors:
from (a): $h$=0.5, (b) $h$=0.56, (c) $h$=0.63, (d) $h$=0.69, (e) $h$=0.75, (f)
h=0.81. Intensities are in arbitrary linear units, the same number of 
detector pixels are shown 
}
\label{fig:truncrod}
\end{figure}

The TR peak has a well identified speckle pattern, as shown in
Figure~\ref{fig:truncrod}. This figure is the result of the scan
along the TR, $h$ varying from 0.5 to 0.81.
This speckle structure is connected to the morphology of the
Si [111] crystalline surface. It is only sensitive to the 
crystalline character of the sample surface, and it is not damped 
by the amorphous surface oxide layer. Between $h$=0.5 (fig. 
\ref{fig:truncrod}\-a) and 
$h$=0.56 (fig. \ref{fig:truncrod}\-b), the coherently irradiated 
region is roughly the same, and we observe that the transverse shape 
of the truncation rod has only small changes.
The $\theta$ angle only varies from 8$^{\circ}$ to 9$^{\circ}$, this 
is a $140 \mu m \times 20\mu m$ irradiated area and the speckle 
structure has slow variations along the [111] direction. The 3\textsc{D} 
pattern observed in this coherent experiment is widely elongated in 
this direction. This indicates that the defects responsible for 
this pattern are surface defects, of atomic scale extension along 
the direction perpendicular to the surface 
\footnote{Here, as we limit our sudy to $h<1$, the (111) atomic layers
of silicon can be assumed as flat layers with a distance of
$a/\sqrt(3)=0.3136$~nm}.  This speckle structure 
is connected to the surface step configuration and it changes with the
sample position. In our case, the regulary spaced steps corresponding 
to the $0.056^{\circ}$ miscut, a period of 0.32~$\mu$m for steps 
of the height of a silicon tetrahedron (0.314~nm), has 
only few irregularities within the few hundreds of steps covered
by the illumination. The 2-3 speckles in the observed area
roughly indicate a 2$\pi$ period error (phase shift fluctuations) 
along this distance. 

\begin{figure}
\begin{tabular}{c}
\includegraphics[clip,width=3.7cm] {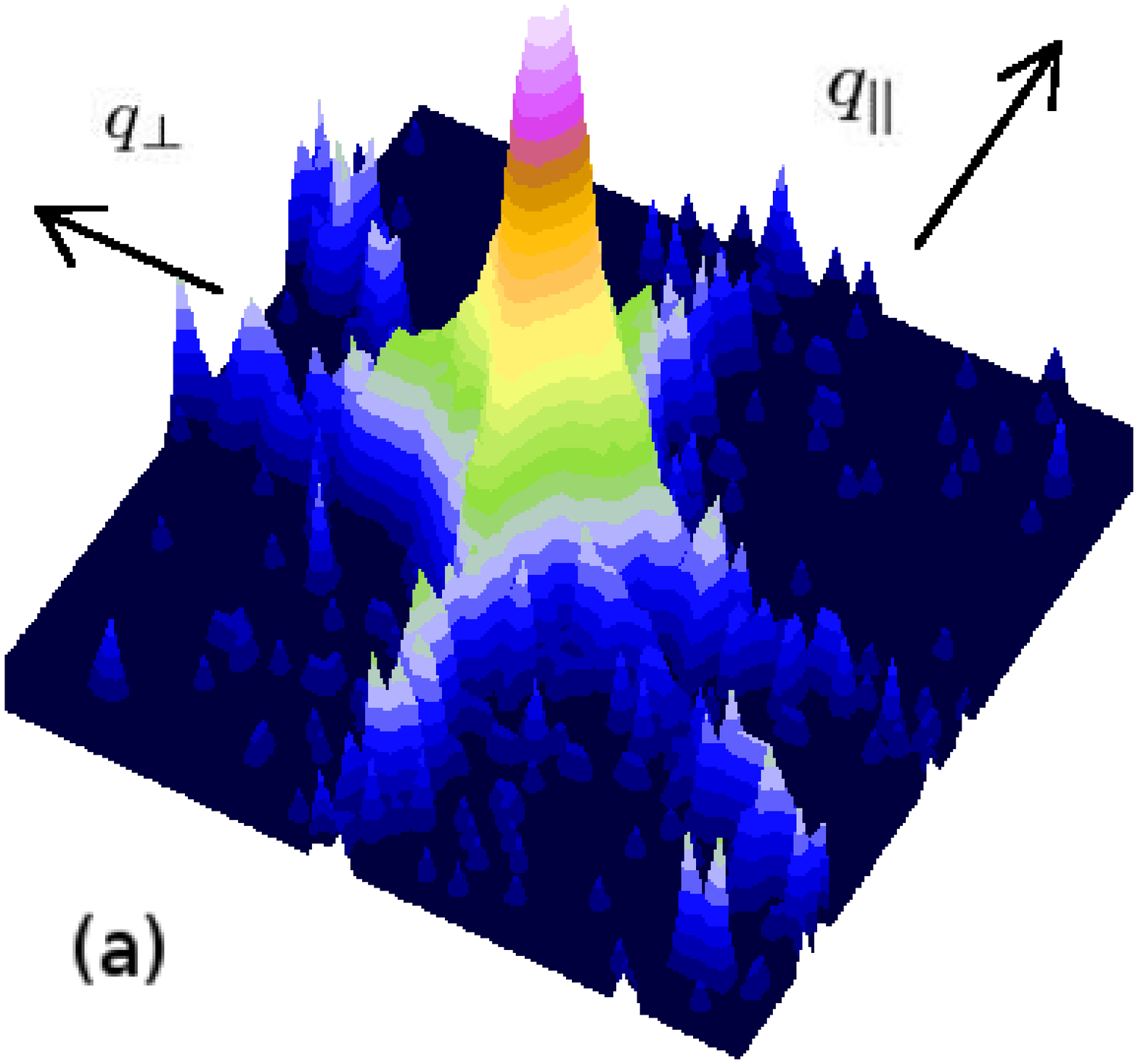}\\
	\includegraphics[clip,width=3.7cm] {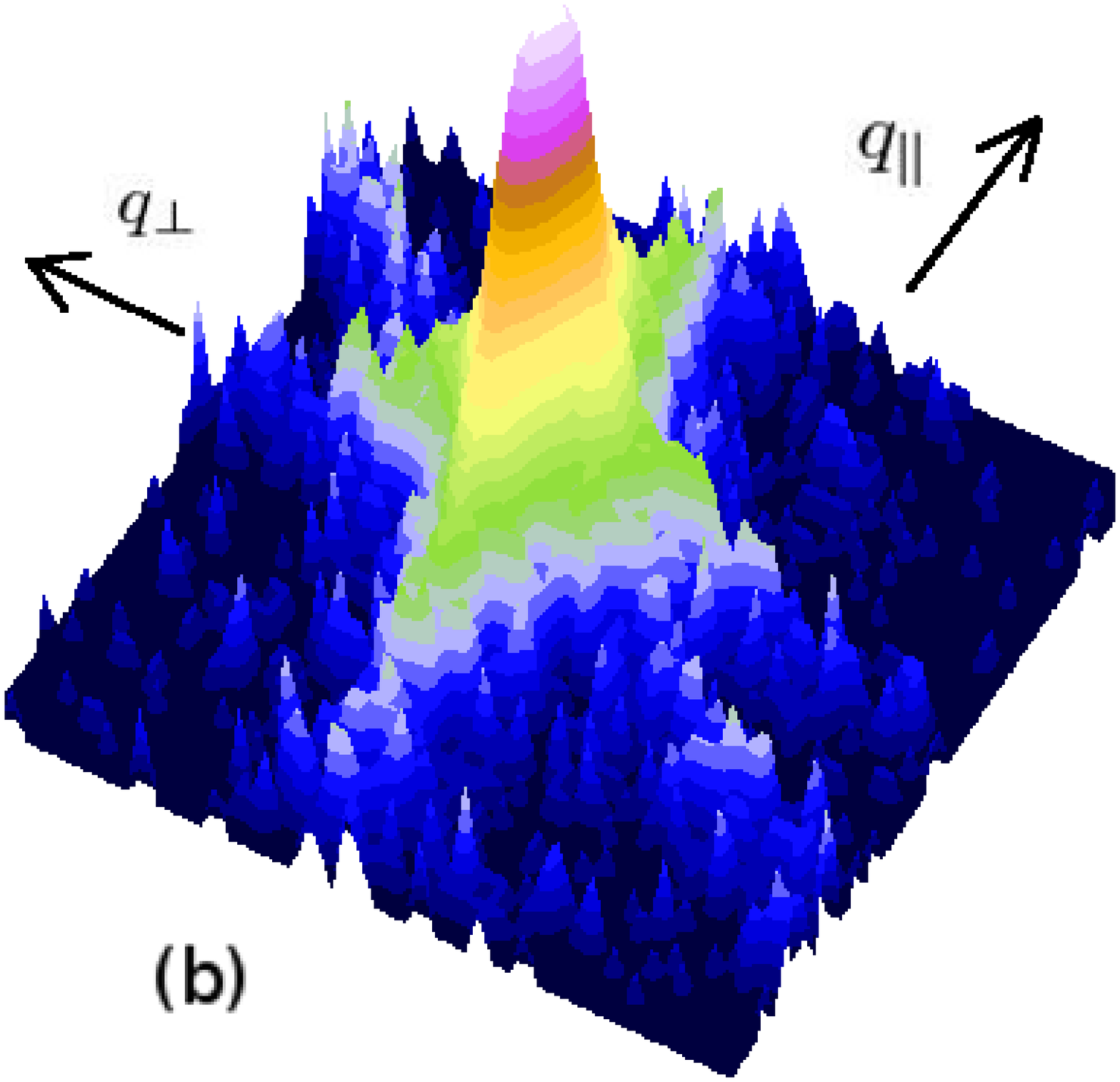}
\end{tabular}
\begin{tabular}{c}
\includegraphics[clip,width=3.8cm] {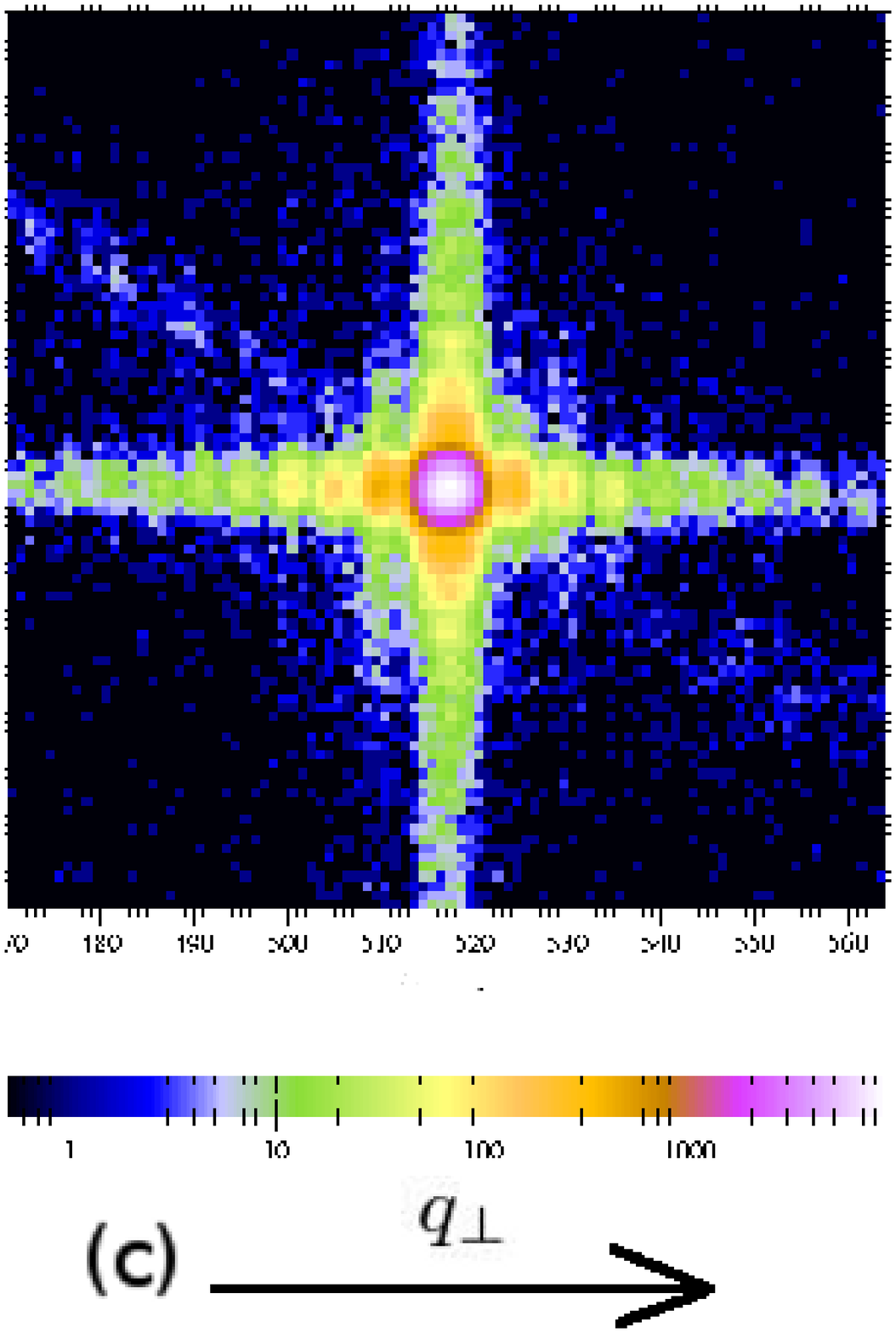}
\end{tabular}
\caption {The shape of the truncation rod in the vicinity of the
Bragg peak (log units): (a) $h$=0.94 and (b) $h$=0.875 in the ($hhh$) direction.
The cross corresponds to the diffraction of the slits limiting 
the beam ($\phi$=20~$\mu$m here). In (c) is shown the scattering
with the aperture $\phi$=10~$\mu$m at $h$=0.99.
}
\label{fig:Bragg}
\end{figure}

For values of ($hhh$) closer to (111) (fig.~\ref{fig:truncrod} d-f),
the pattern is reduced essentially to a single peak. In this case,
the small value of the difference $|\Delta\vec{Q}|$ between the Bragg
peak and ($hhh$) makes the observation of the atomic scale 
surface defects difficult. The scattering corresponds to perfect
atomic planes in the sample. In this case, diffraction does not 
significanly change the shape of the wave front of the incoming 
beam. This is shown in Fig~\ref{fig:Bragg}, where the cross-shaped
diffracted beam reflects the slits selecting the coherent beam. In 
fig.~\ref{fig:Bragg}, a diffuse streak is also observed, with a nearly 
30$^{\circ}$ angle with the horizontal axis q$_{\perp}$, which we 
ascribe to residual stress of the crystal surface connected to polishing.

These results were obtained from a silicon single crystal which is
a model system to discuss the use of the coherent scattering 
technique for the study of crystal surfaces. Can this technique
be used for surface studies in other crystalline systems? In the 
case of metals, like copper, lead or gold, crystals are highly 
distorted and any mechanical polishing creates a large number of 
dislocations in the vicinity of the surface. The observation of a 
coherently diffracting truncation rod needs a high crystalline quality 
in the irradiated region, otherwise the mosaic structure inhibits 
the observation of speckles connected to
surface irregularities. For metals, ``in situ'' restoration at high 
enough temperature can provide subgrains of a few micron size. 
As it seems difficult to obtain dislocation-free regions larger
than this, the size of the coherent beam should also be 
in the (sub)micrometer range. Beams of that size are now 
available with the modern X-ray focusing techniques (refracting 
lenses, Fresnel lenses or mirrors) and coherent beams can be 
selected by simply closing slits at the entrance of the focusing 
setup. 
From Eq.~\ref{equ:coh_vert} and \ref{equ:coh_hor}, small values of 
$\phi$, comparable with $\Lambda_l$, open the possibility of 
coherent measurements at large asymmetric angles.

One important problem here is the low intensity at the AB position. 
In this experiment, the measured intensity was of the order of 10~ph/pixel
for 2,500~s. This intensity has a $Z^2$ dependence and ``heavy'' metals can 
provide some improvements. Focusing also limits the number of steps 
and other defects in the irradiated area. This reduces the number of 
speckles that share the AB intensity, making easier their observation. In 
this case, the technique is limited to small values of $|\Delta\vec(Q)|$ 
and only large scale step movements can be observed (some tens of 
nanometers). Optics improvements and the use of x-ray sources with higher 
brilliance (NSLS II , Petra III or even free-electron lasers) will raise 
this intensity of some orders of magnitude, reducing the accessible time 
scale.

\begin{acknowledgments}
The authors are grateful to Benoit Picut from the ESRF optics laboratory
who has provided a high quality mechano-chemically polished silicon crystal.
\end{acknowledgments}

\bibliographystyle{apsrev}
\bibliography{pap}



\end{document}